\documentclass[10pt]{article}
\usepackage{graphicx}

\def\Title#1{\begin{center} {\Large #1 } \end{center}}
\def\Author#1{\begin{center}{ \sc #1} \end{center}}
\def\Address#1{\begin{center}{ \it #1} \end{center}}

\newcommand\pubblock{\rightline{\begin{tabular}{l} Proceedings of the Fifth Annual LHCP\\ \pubnumber\\
         \pubdate  \end{tabular}}}

\newenvironment{Abstract}{\begin{quotation} \begin{center} 
             \large ABSTRACT \end{center}\bigskip 
      \begin{center}\begin{large}}{\end{large}\end{center} \end{quotation}}

\newenvironment{Presented}{\begin{quotation} \begin{center} 
             PRESENTED AT\end{center}\bigskip 
      \begin{center}\begin{large}}{\end{large}\end{center} \end{quotation}}

\def\beq{\begin{equation}}
\def\eeq#1{\label{#1}\end{equation}}
\def\eeqn{\end{equation}}

\def\beqa{\begin{eqnarray}}
\def\eeqa#1{\label{#1}\end{eqnarray}}
\def\eeqan{\end{eqnarray}}

\let\bar=\overbar

\def\Dslash{\not{\hbox{\kern-4pt $D$}}}
\def\dslash{\not{\hbox{\kern-2pt $\del$}}}

\def\msb{{\bar{\ssstyle M \kern -1pt S}}}

\usepackage{color}

\newcommand\pubnumber{CMS-CR-2017/244}

\newcommand\pubdate{\today}

\def\affiliation{
On behalf of the ATLAS and CMS Collaborations, \\
Instituto de F\'isica de Cantabria \\
(CSIC - Universidad de Cantabria), Spain}

\begin{document}

\large
\begin{titlepage}
\pubblock

\vfill
\Title{New results on Higgs boson properties}
\vfill

\Author{J\'onatan Piedra}
\Address{\affiliation}
\vfill
\begin{Abstract}
We present the latest ATLAS and CMS measurements of several properties of the
Higgs boson, such as signal-strength modifiers for the main production modes,
fiducial and differential cross sections, and the Higgs mass. We have analyzed
the 13~TeV proton-proton LHC collision data recorded in 2016, corresponding to
integrated luminosities up to 36.1~${\rm fb}^{-1}$. Results for the
${\rm H\to ZZ}\to 4\ell$ (${\rm \ell = e\mu}$),
${\rm H}\to\gamma\gamma$, and
${\rm H}\to\tau\tau$
decay channels are presented. In addition, searches for new phenomena in the
${\rm H}\to\gamma\gamma + E_{\rm T}^{\rm miss}$ and
${\rm H}\to{\rm b\bar{b}} + E_{\rm T}^{\rm miss}$ decay channels are presented.
\end{Abstract}
\vfill

\begin{Presented}
The Fifth Annual Conference\\
 on Large Hadron Collider Physics \\
Shanghai Jiao Tong University, Shanghai, China\\ 
May 15-20, 2017
\end{Presented}
\vfill
\end{titlepage}
\def\thefootnote{\fnsymbol{footnote}}
\setcounter{footnote}{0}
%

\normalsize 


\section{Introduction}

The discovery of the Higgs boson was announced in
2012~\cite{Aad:2012tfa,Chatrchyan:2012ufa,Chatrchyan:2013lba} by the
ATLAS~\cite{Aad:2008zzm} and CMS~\cite{Chatrchyan:2008aa} collaborations based
on proton-proton
collisions collected at the CERN LHC at the centre of mass energies of 7 and
8~TeV. Since then a huge effort has been made in the determination of the
properties of this newly found particle. The dataset already collected at 13~TeV
allows inclusive Higgs boson measurements to be repeated. Furthermore, the
increased centre-of-mass energy results in much larger cross sections for events
at high partonic centre-of-mass energy. This implies improved sensitivity to a
variety of interesting physics processes, such as Higgs bosons produced at high
transverse momentum.

In this document we present the latest ATLAS and CMS measurements of several
properties of the Higgs boson in different decay channels, such as
${\rm H\to ZZ}$, ${\rm H\to\gamma\gamma}$ and ${\rm H\to\tau\tau}$. In addition,
we also present results on searches for phenomena beyond the Standard Model, in
Higgs decays to $\gamma\gamma$ or ${\rm b\bar{b}}$, with $E_{\rm T}^{\rm miss}$
in the final state.

\section{\boldmath ${\rm H}\to{\rm ZZ}$}

The ${\rm H\to ZZ\to 4\ell}$ decay channel (${\rm \ell = e,\mu}$) has a large
signal-to-background ratio due to the complete reconstruction of the final state
decay products and excellent lepton momentum resolution, making it one of the
most important channels for studies of the Higgs boson's properties. Here we
present measurements of properties of the Higgs boson in this channel at 13~TeV,
for both the ATLAS and CMS collaborations~\cite{Aaboud:2017oem,Sirunyan:2017exp}.

\begin{figure}[htb]
\centering
\includegraphics[height=2in]{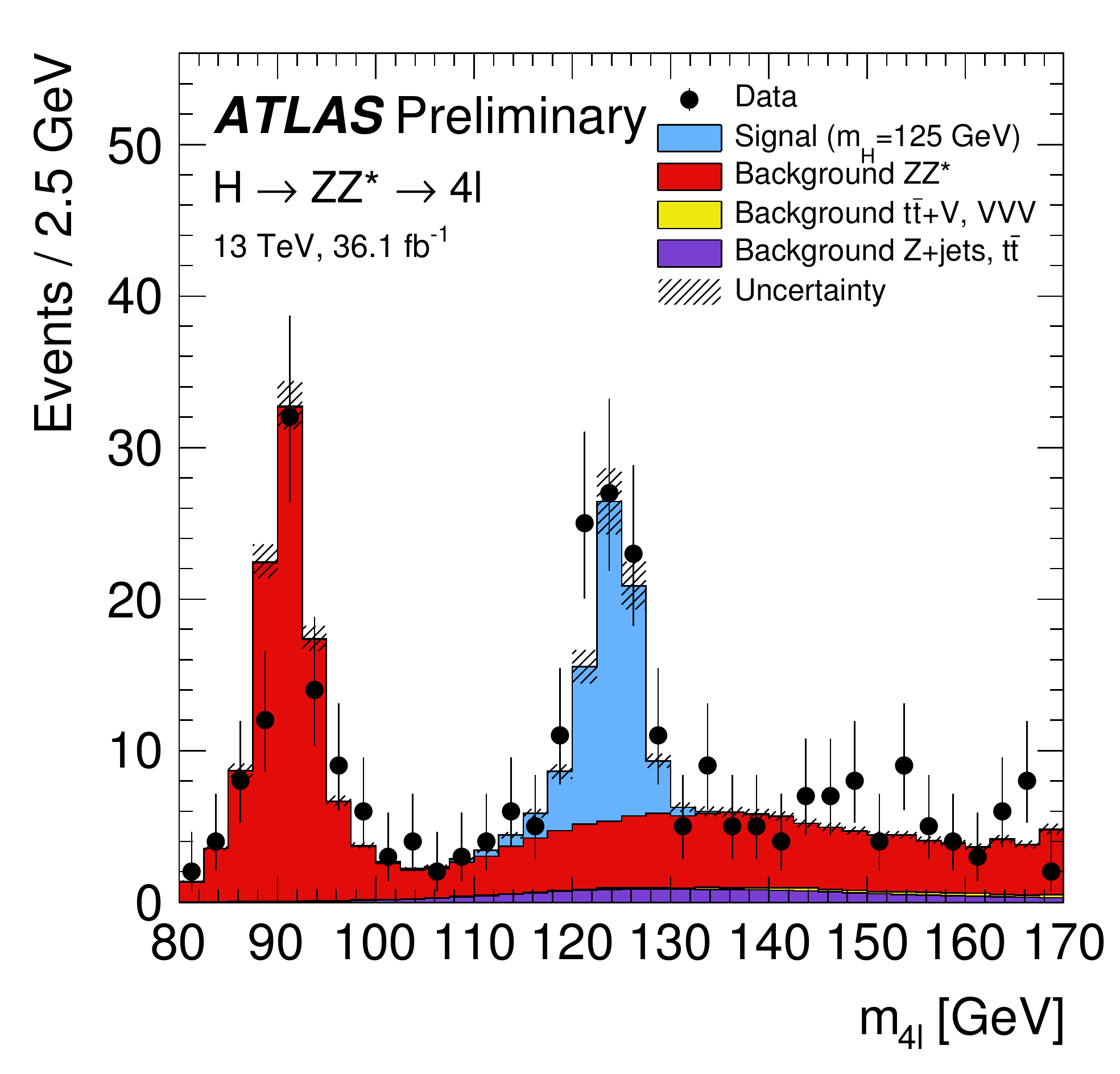}
\includegraphics[height=2in]{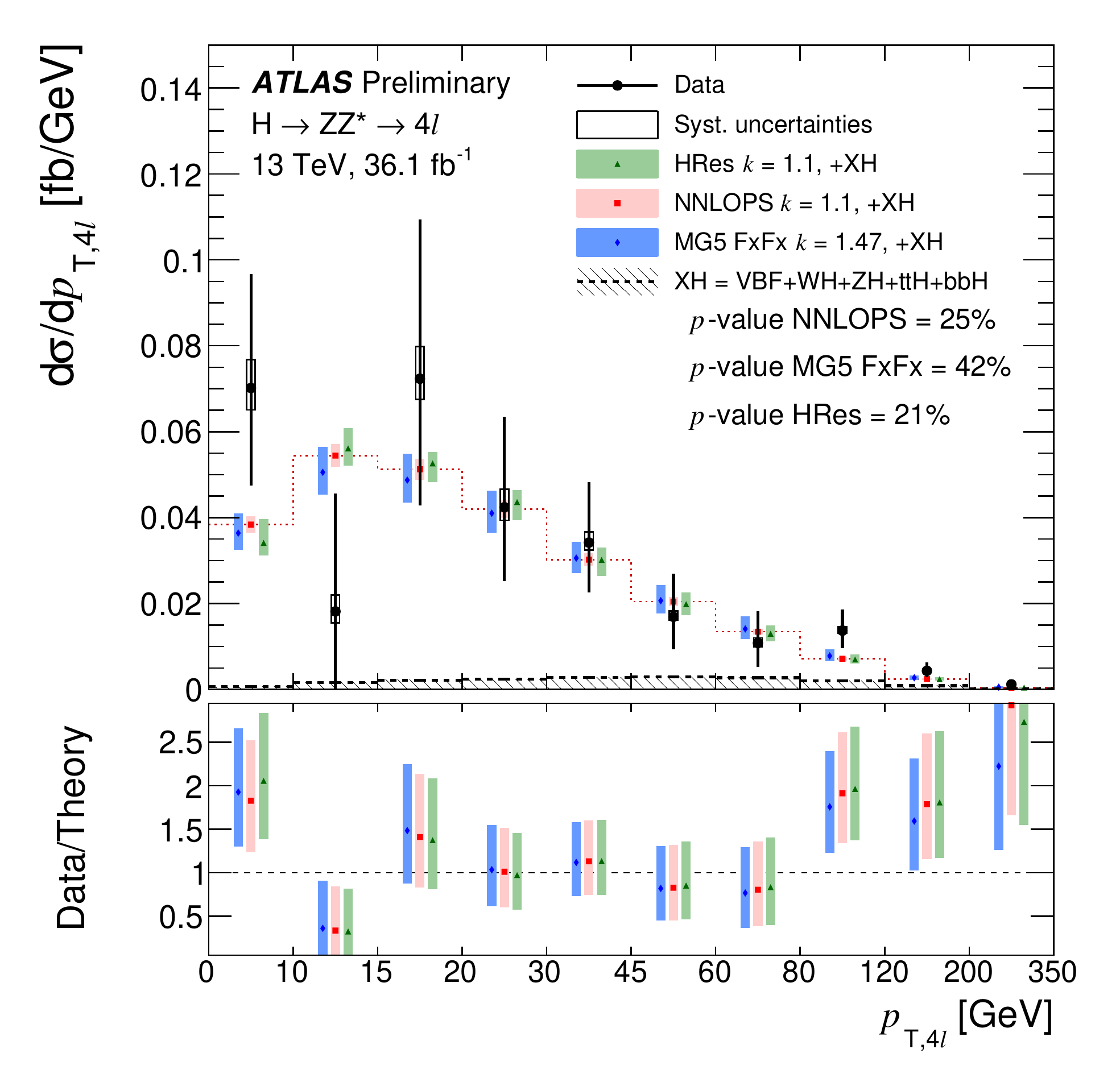}
\includegraphics[height=2in]{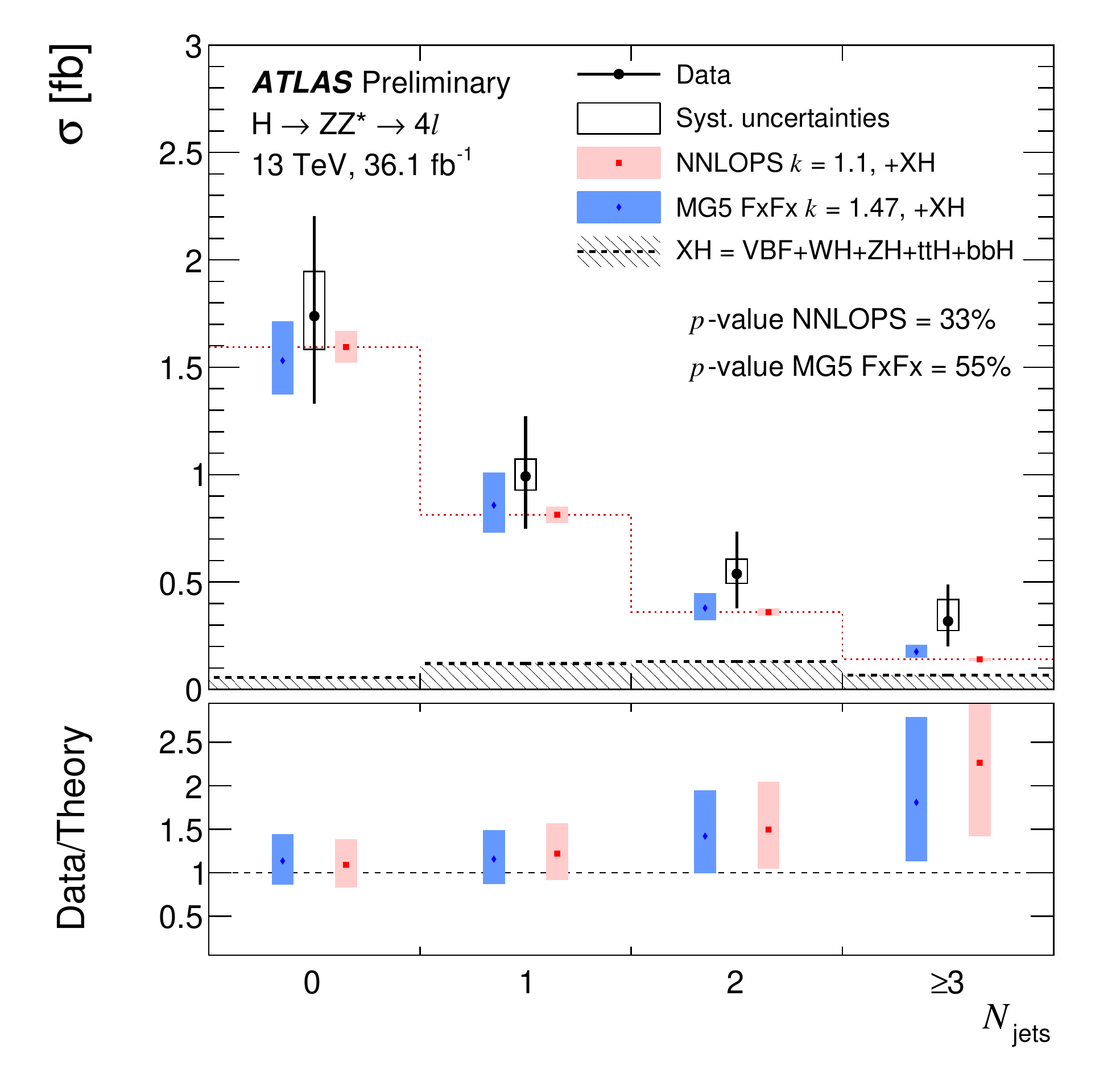}\\
\includegraphics[height=2in]{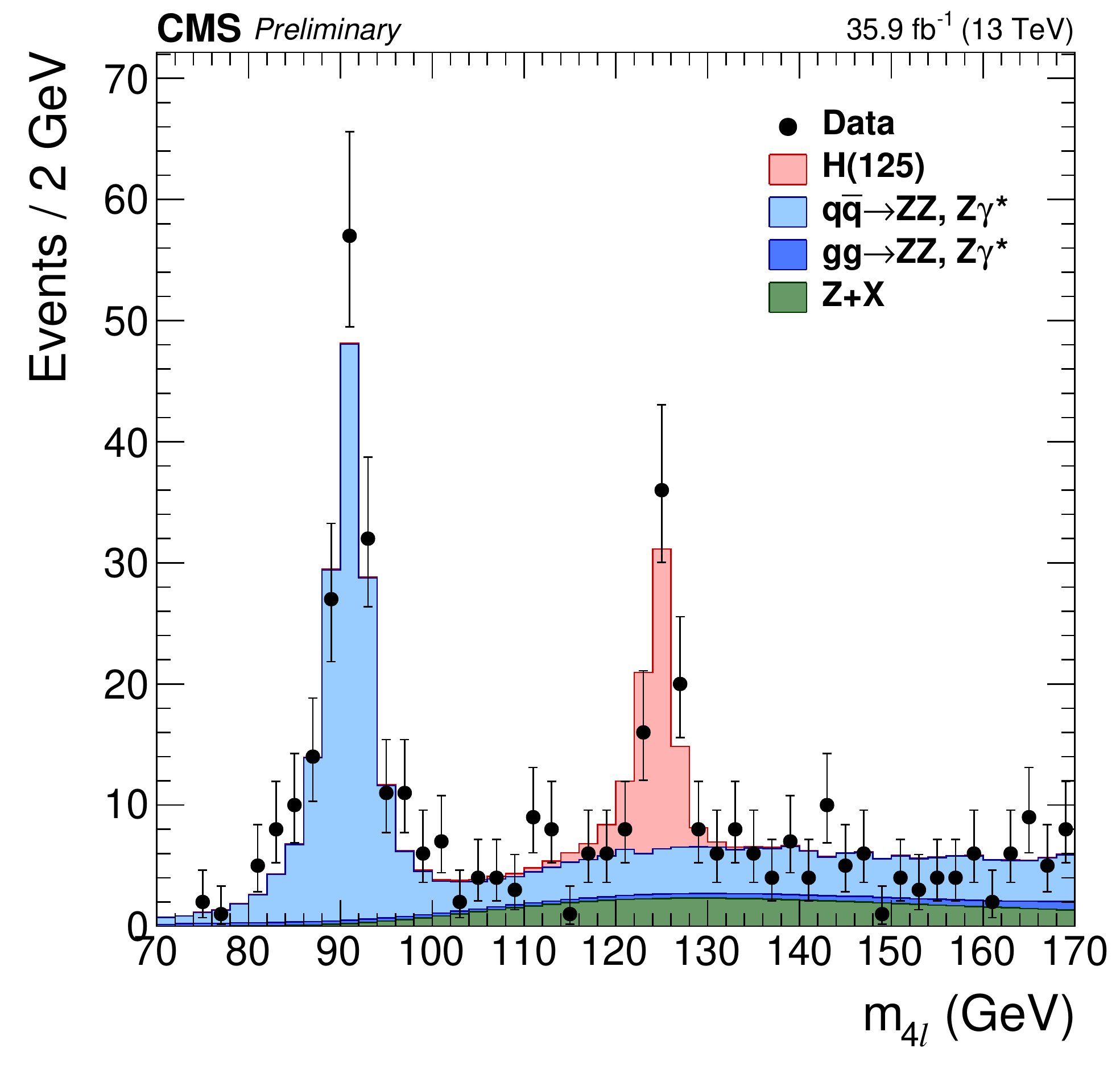}
\includegraphics[height=2in]{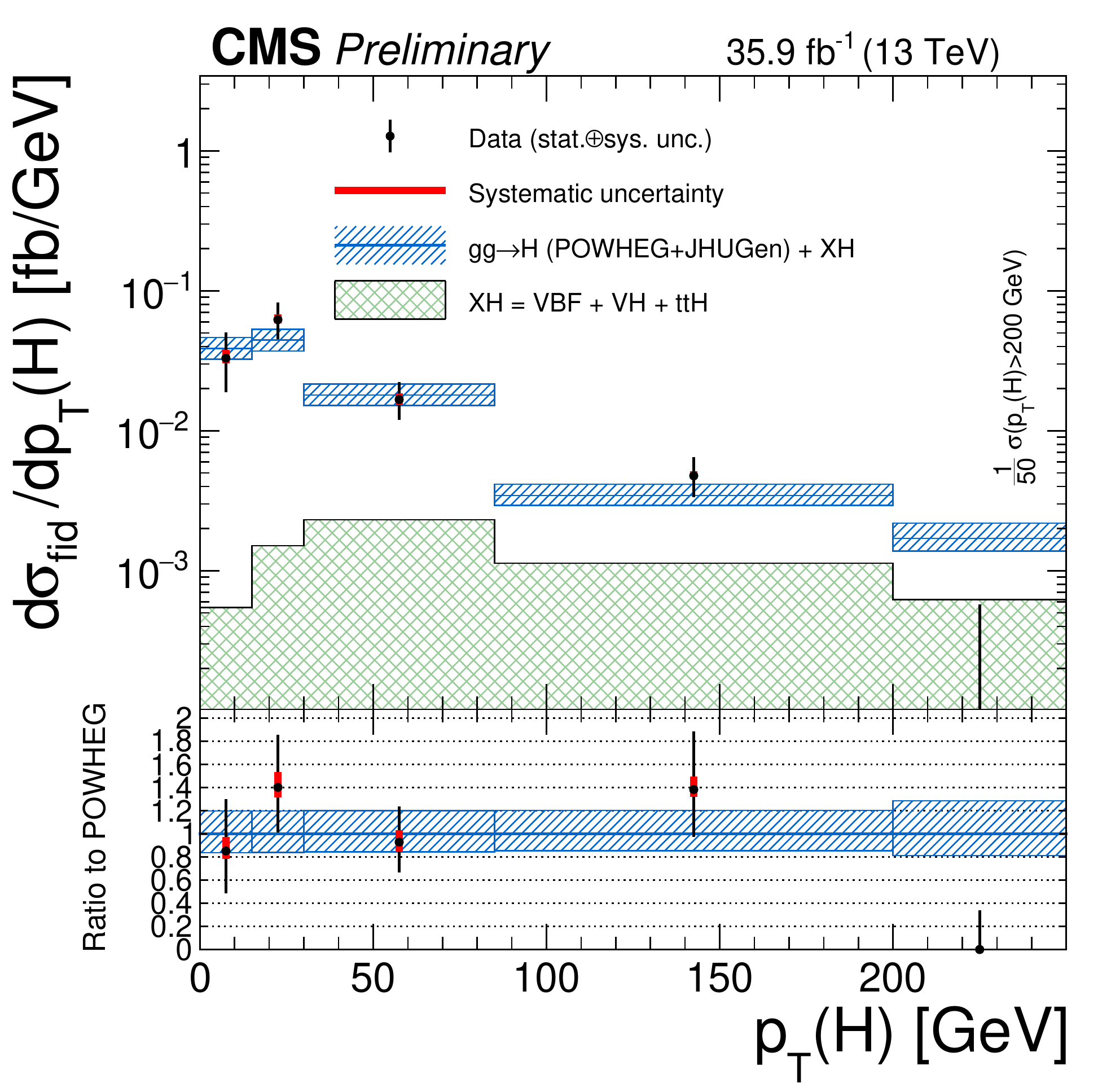}
\includegraphics[height=2in]{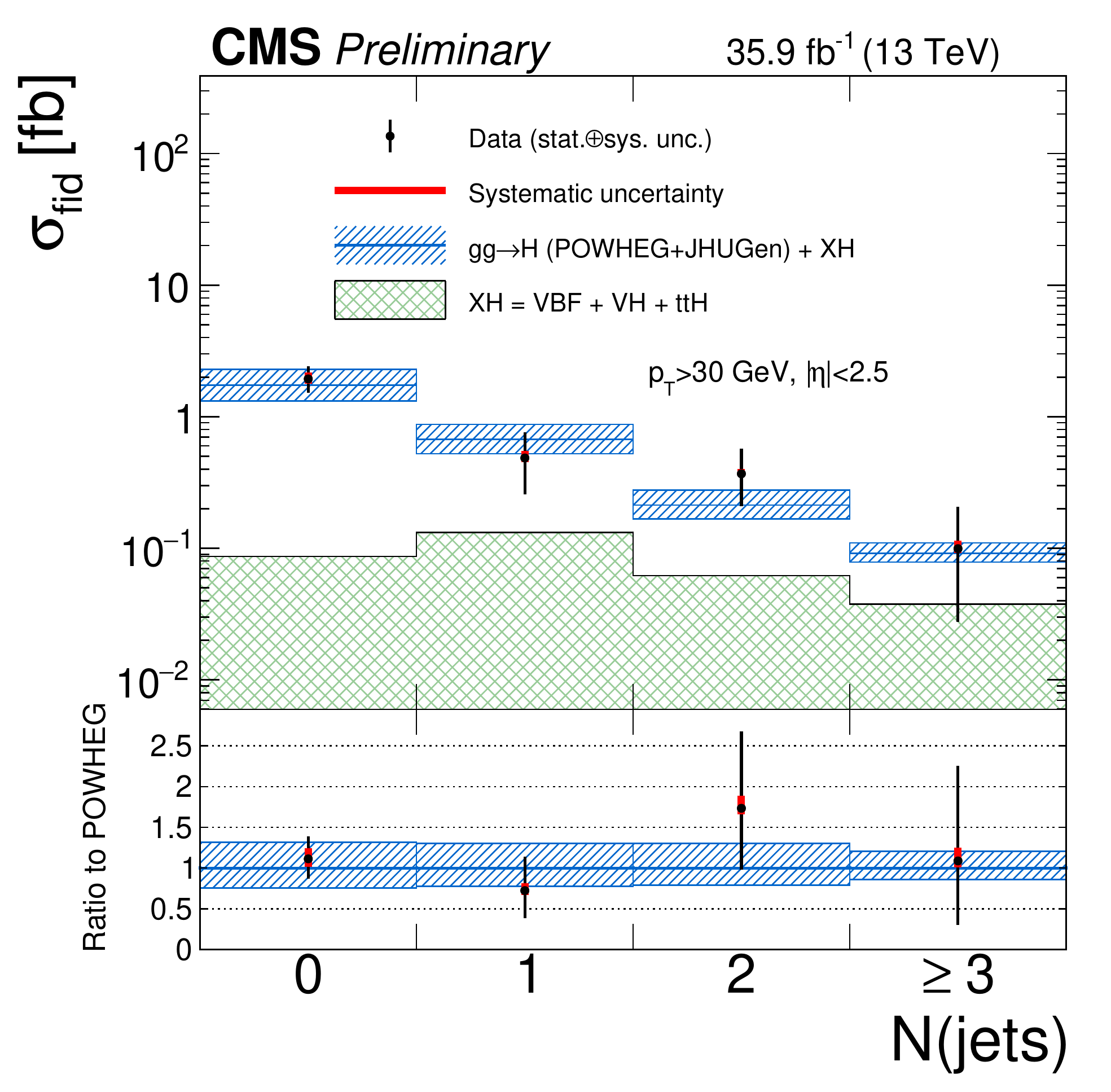}
\caption{
  (Top left) ATLAS four-lepton invariant mass distribution of the selected
  events~\cite{Aaboud:2017oem}. The systematic uncertainty on the prediction is shown
  by the dashed band. (Top center and right) ATLAS differential fiducial cross
  sections~\cite{Aaboud:2017oem}, for the transverse momentum of the Higgs boson and
  the number of jets.
  The measured cross sections are compared to different ggH predictions, and
  predictions for all other Higgs boson production modes XH are added.
  (Bottom left) CMS four-lepton invariant mass distribution of the selected
  events~\cite{Sirunyan:2017exp}.
  (Bottom center and right) CMS differential fiducial cross sections, for the
  transverse momentum of the Higgs boson and the number of jets~\cite{Sirunyan:2017exp}.
  The sub-dominant component of the signal (VBF + VH + ttH) is denoted as XH.
}
\label{fig:figure-ZZ}
\end{figure}

\begin{figure}[htb]
\centering
\includegraphics[height=2in]{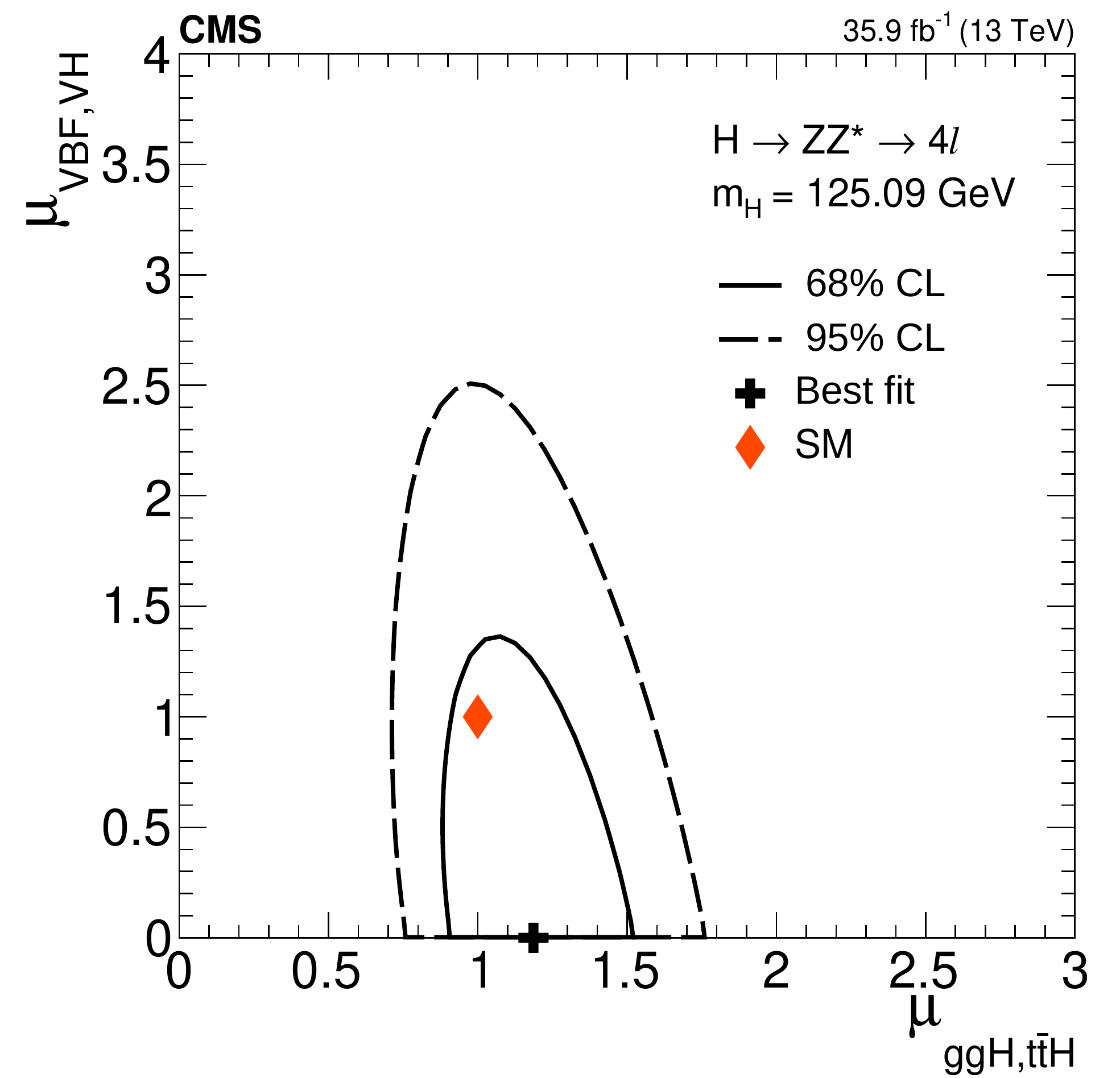}
\includegraphics[height=2in]{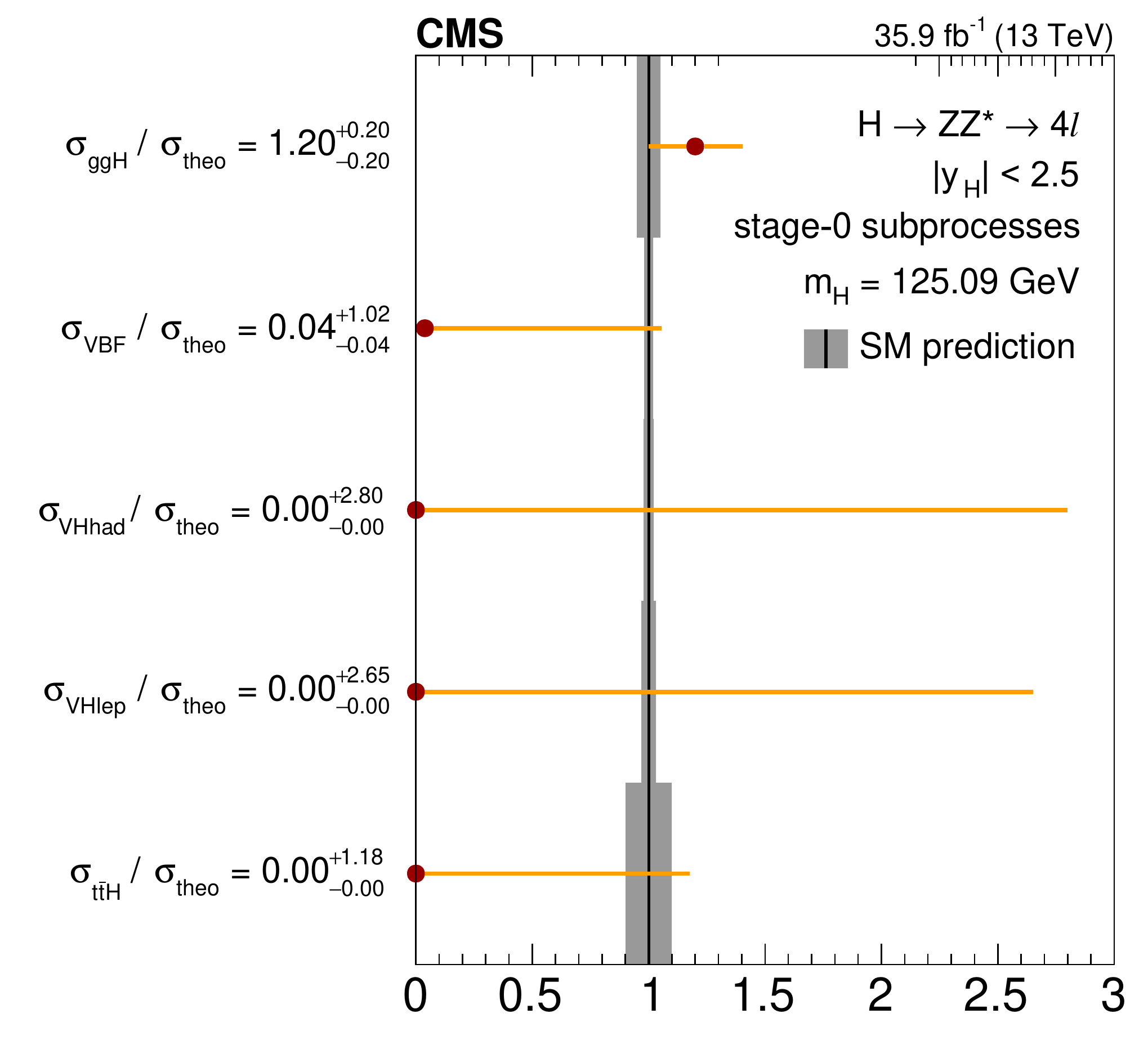}
\includegraphics[height=2in]{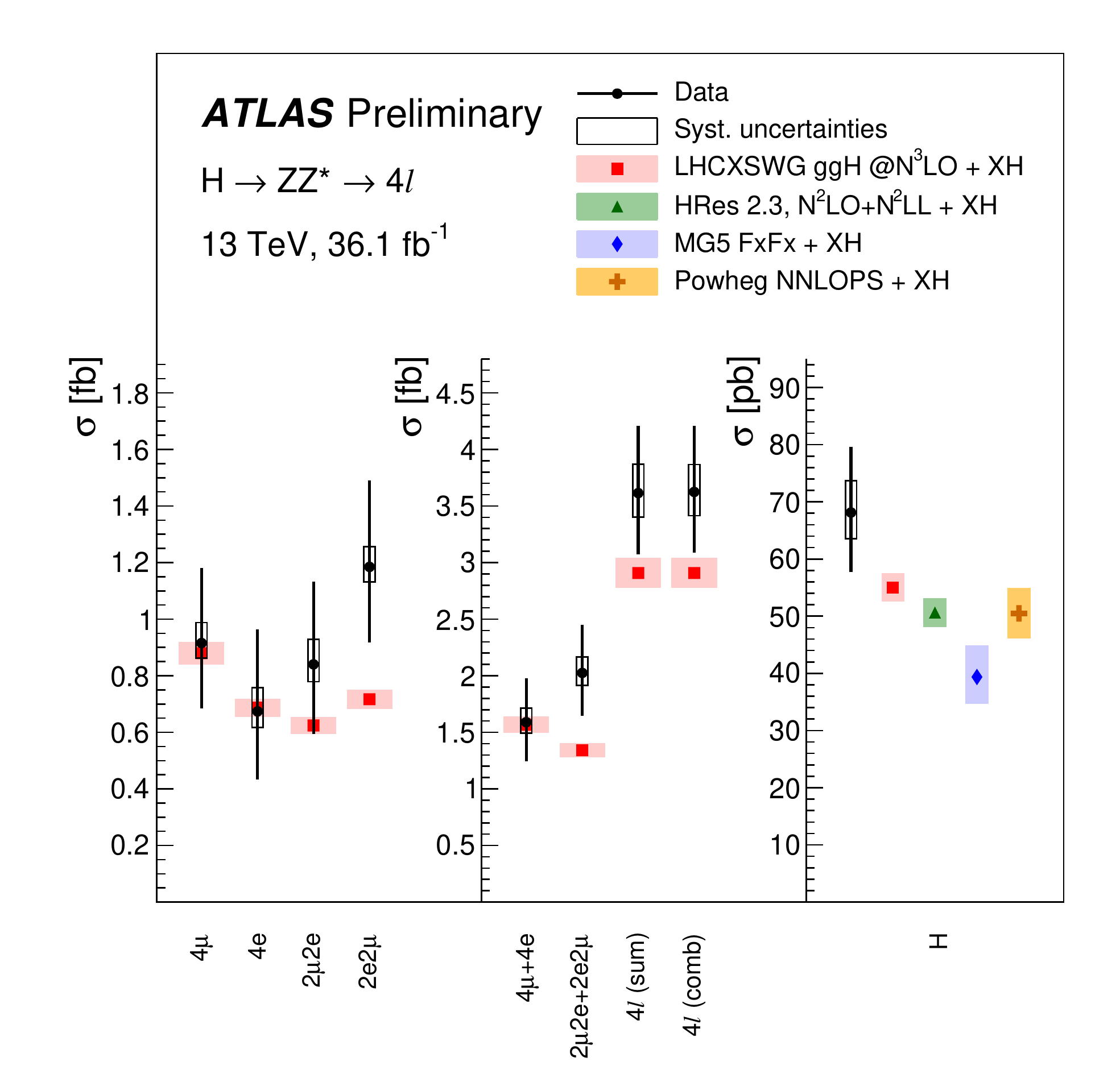}
\caption{
  (Left) Result of the 2D likelihood scan for the ${\rm \mu_{ggH,ttH}}$
  and ${\rm \mu_{VBF,VH}}$ signal-strength modifiers~\cite{Sirunyan:2017exp}.
  The solid and dashed
  contours show the 68\% and 95\% CL regions, respectively. The cross
  indicates the best-fit values, and the diamond represents the expected
  values for the SM Higgs boson.
  (Center) Results of the fit for simplified template cross sections for
  the stage 0 sub-processes, normalized to the SM prediction~\cite{Sirunyan:2017exp}.
  (Right) The ATLAS fiducial cross sections and total cross
  section of Higgs-boson production measured in the $4\ell$
  final state~\cite{Aaboud:2017oem}. The SM prediction includes the N3LO
  calculation for ggH production.
}
\label{fig:figure-ZZ-2}
\end{figure}

In the CMS analysis~\cite{Sirunyan:2017exp} the full kinematic information from each
event using either
the Higgs boson decay products or associated particles in its production is
extracted with matrix element calculations, and used to form several kinematic
discriminants. Both H boson decay kinematics and kinematics of associated
production of H+jet, H+2 jets, VBF, ZH, and WH are explored. In order to improve
the sensitivity to the Higgs boson production mechanisms, the selected events
are classified into mutually exclusive categories, based on jet multiplicity,
b-tagged jets, the existence of additional leptons, and the aforementioned
kinematic discriminants. The reconstructed four-lepton invariant mass distribution
is shown in Figure~\ref{fig:figure-ZZ} (bottom left) for the sum of the 4e,
4$\mu$ and 2e2$\mu$ subchannels. To extract the signal strength for the excess
of events observed in the Higgs boson peak region a multi-dimensional fit
is performed, on the four-lepton invariant mass and one of the kinematic
discriminants. The fit is performed for several signal-strength modifiers
controlling the contribution of the main SM Higgs boson production modes: ggH,
VBF, WH, ZH and ${\rm ttH}$, getting for all of them values consistent
with the SM expectaction. Cross section ratios for the Stage 0 STXS can be seen
in Figure~\ref{fig:figure-ZZ-2} (center). The inclusive observed signal-strength
modifier is ${\rm 1.05_{-0.14}^{+0.15}(stat.)_{-0.09}^{+0.11}(syst.)}$.
Two signal-strength modifiers ${\rm \mu_{ggH,ttH}}$ and ${\rm \mu_{VBF,VH}}$ are
introduced as scale factors for the fermion and vector-boson induced contributions
to the expected SM cross section. A two-dimensional fit is performed assuming a
mass of ${\rm m_H = 125.09~GeV}$ leading to the measurements of
${\rm \mu_{ggH,ttH} = 1.20_{-0.31}^{+0.35}}$ and
${\rm \mu_{VBF,VH} = 0.00_{-0.00}^{+1.37}}$. The 68\% and 95\% CL contours are
shown in Figure~\ref{fig:figure-ZZ-2} (left). We have also
measured the cross section for the production and decay ${\rm pp \to H \to 4\ell}$
within a fiducial volume defined to closely match the reconstruction selection.
The integrated fiducial cross section is measured to be
${\rm \sigma_{fid} = 2.90_{-0.44}^{+0.48}(stat.)_{-0.22}^{+0.27}(syst.)~fb}$.
This can be compared to the SM expectation of
${\rm 2.72 \pm 0.14~fb}$. The measured differential cross section results for
${\rm p_{T}^{H}}$ and ${\rm N_{jets}}$ can also be seen in Figure~\ref{fig:figure-ZZ}
(bottom center and bottom right). The dominant sources of systematic uncertainty are
the experimental uncertainties in the lepton identification efficiencies and
luminosity measurement. In order to improve the four-lepton invariant mass
resolution a 3D kinematic fit is performed using a mass constraint on the
intermediate Z resonance. The simultaneous fit on the four-lepton
invariant mass, the mass resolution and one of the kinematic discriminants
gives ${\rm m_{H} = 125.26 \pm 0.20(stat.) \pm 0.08(syst.)~GeV}$. The width of
the Higgs boson is also measured in the range ${\rm 105 < m_{4\ell} < 140~GeV}$.
No assumptions need to be made about the presence of BSM particles or
interactions which could affect the Higgs boson couplings either in production
or in decay. We obtain a width that is constrained to be
${\rm \Gamma_{H} < 1.10~GeV}$ at 95\% CL.

The ATLAS~\cite{Aaboud:2017oem} four-lepton invariant mass can be seen in
Figure~\ref{fig:figure-ZZ} (top left). Similarly to the CMS analysis, fiducial
cross sections are defined at particle-level, with a selection chosen
to closely match the detector-level analysis. The inclusive fiducial cross
sections of Higgs-boson production at 13 TeV measured using
${\rm H \to ZZ \to 4\ell}$ decays are presented in Figure~\ref{fig:figure-ZZ-2}
(right). The fiducial cross section is measured
to be ${\rm 3.62_{-0.50}^{+0.53}(stat.)_{-0.20}^{+0.25}(syst.)~fb}$, in agreement
with the SM prediction of ${\rm 2.91 \pm 0.13~fb}$. The measured total cross section is
${\rm 69_{-9}^{+10}(stat.)_{-4}^{+5}(syst.)~pb}$, to be compared with the
SM prediction of ${\rm 55.6 \pm 2.5~pb}$. The measured differential cross sections
for ${\rm p_T^{H}}$ and ${\rm N_{jets}}$, together with their comparisons to SM
predictions, are presented in Figure~\ref{fig:figure-ZZ} (top center and right).

\section{\boldmath ${\rm H}\to\gamma\gamma$}

The Higgs boson decay into two photons (${\rm H\to\gamma\gamma}$) is a
particularly attractive way to study the properties of the Higgs boson. Despite
the small branching ratio, a reasonably large signal yield can be obtained thanks
to the high photon reconstruction identification efficiencies at both ATLAS and
CMS. Furthermore, the signal manifests itself as a narrow peak in the diphoton
invariant mass spectrum on top of a smoothly falling background, and the Higgs
boson signal yield can be measured using an appropriate fit.

\begin{figure}[htb]
\centering
\includegraphics[height=2in]{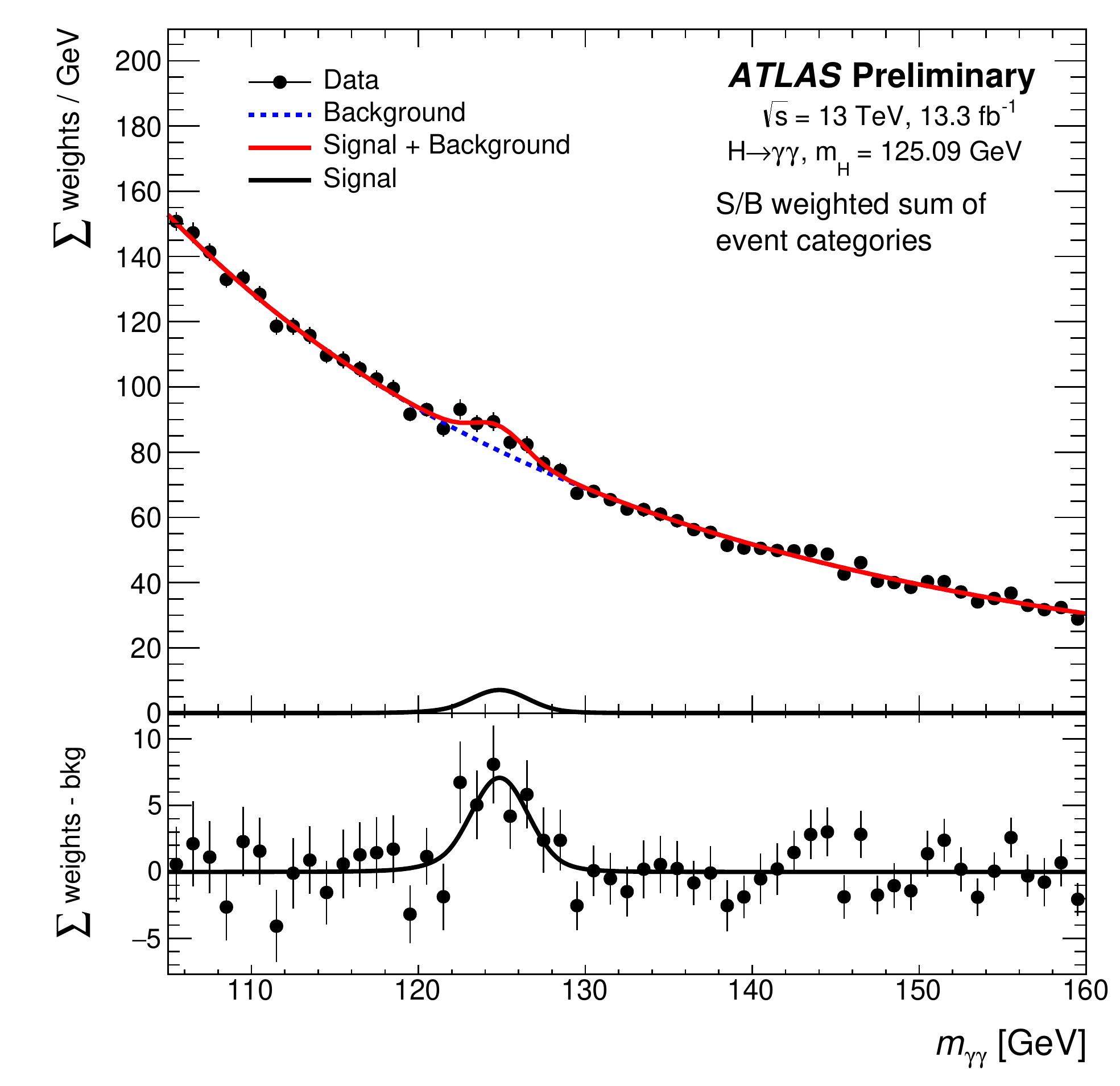}
\includegraphics[height=2in]{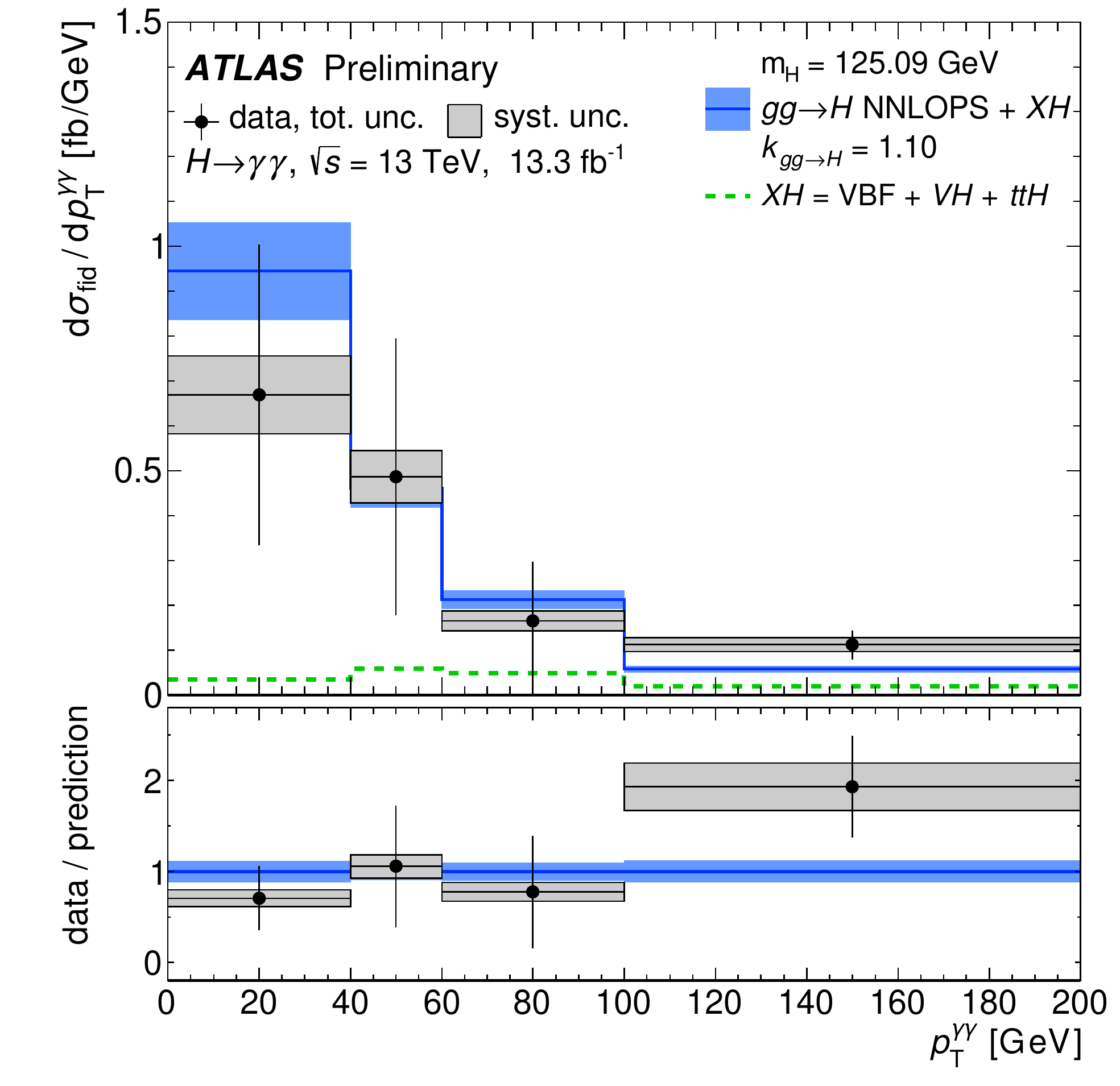}
\includegraphics[height=2in]{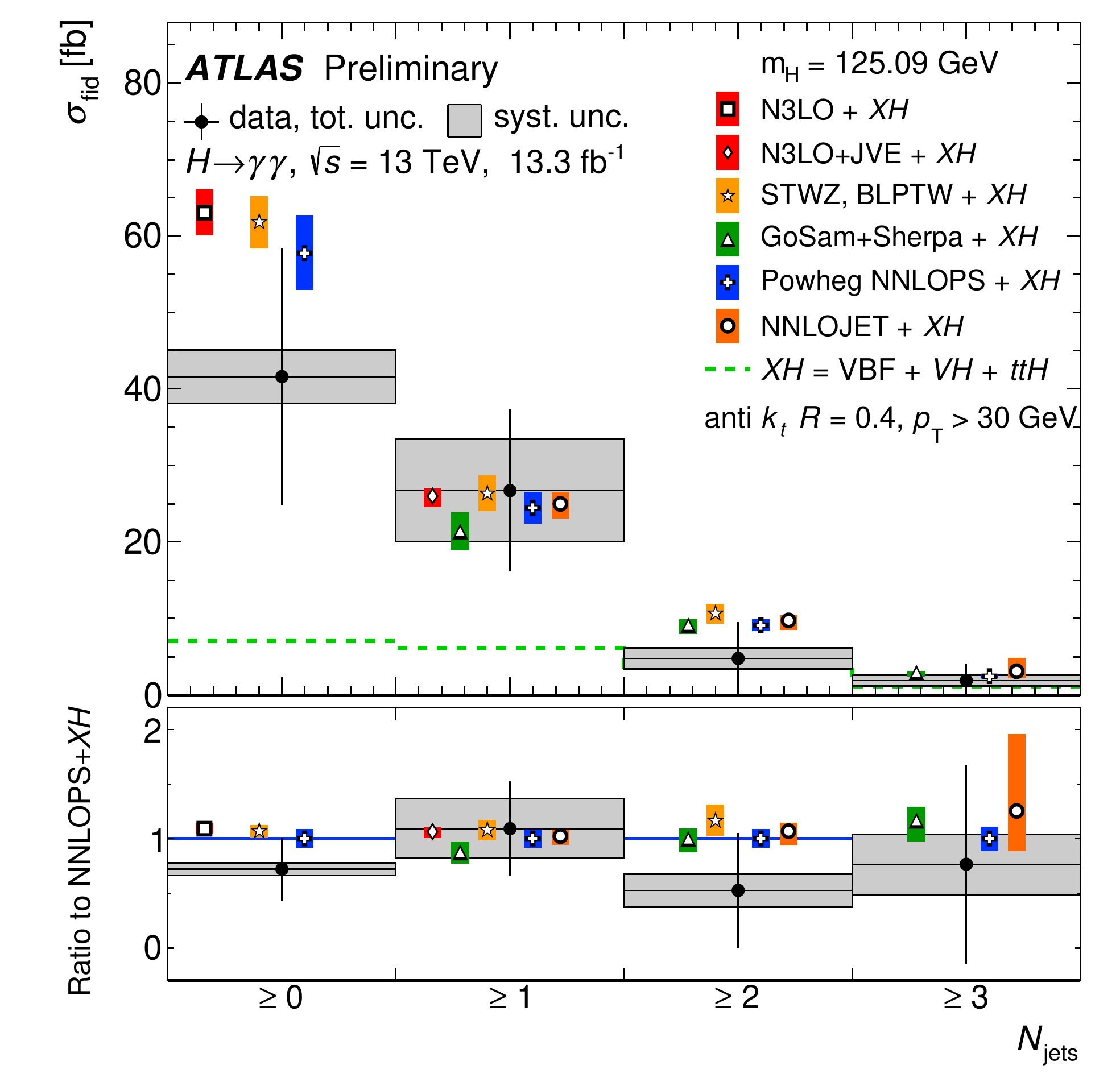}\\
\includegraphics[height=2in]{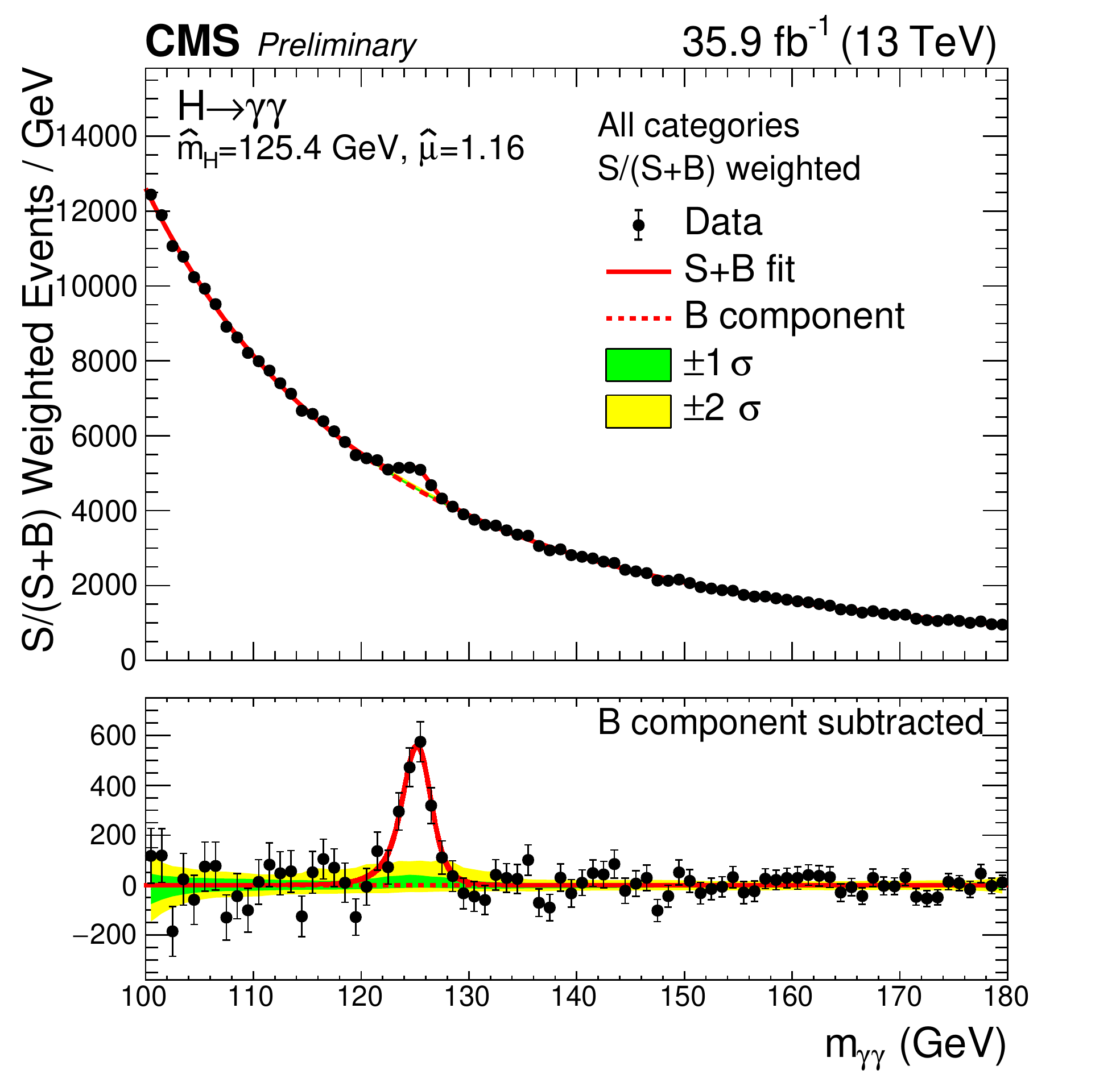}
\includegraphics[height=2in]{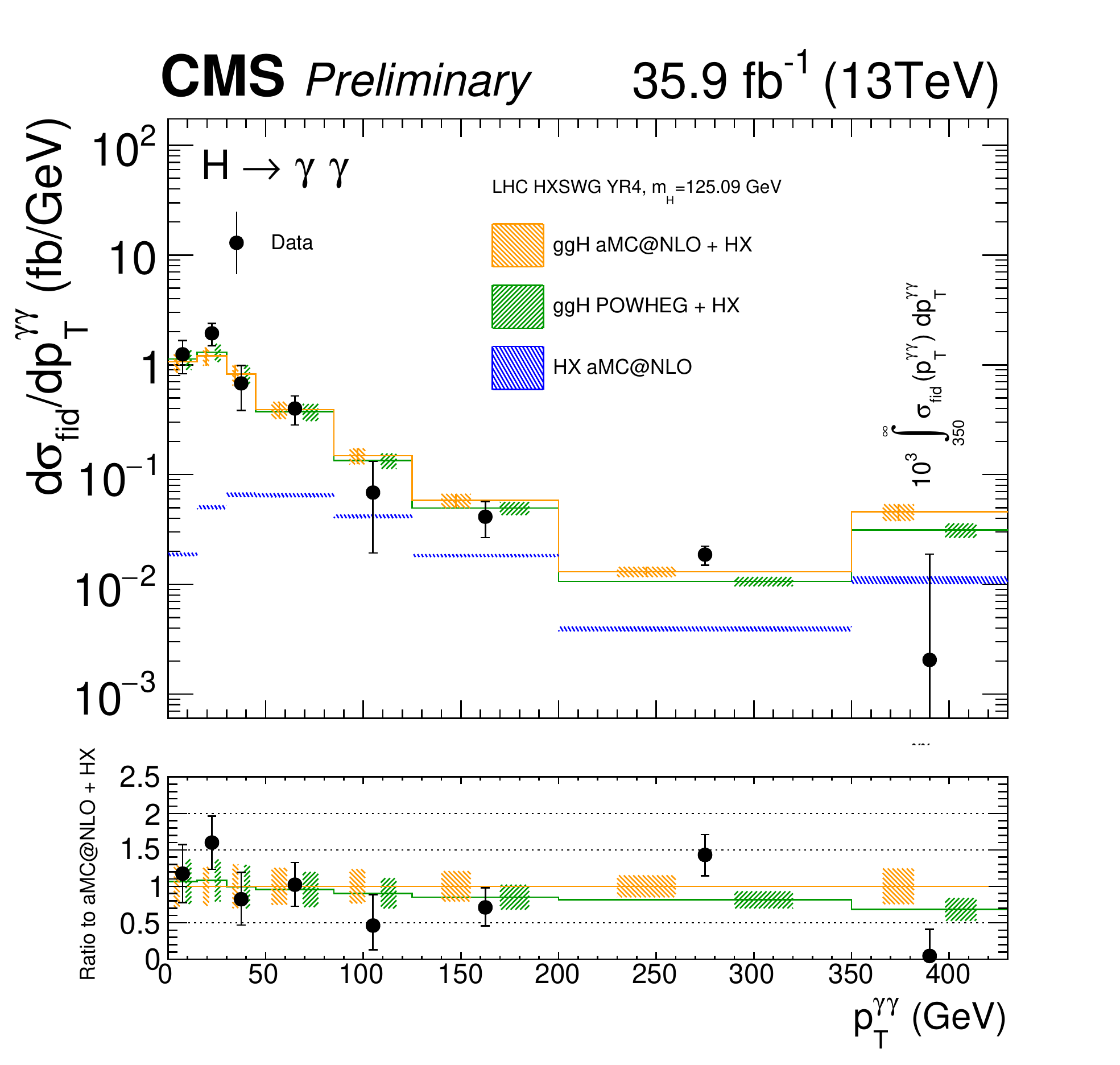}
\includegraphics[height=2in]{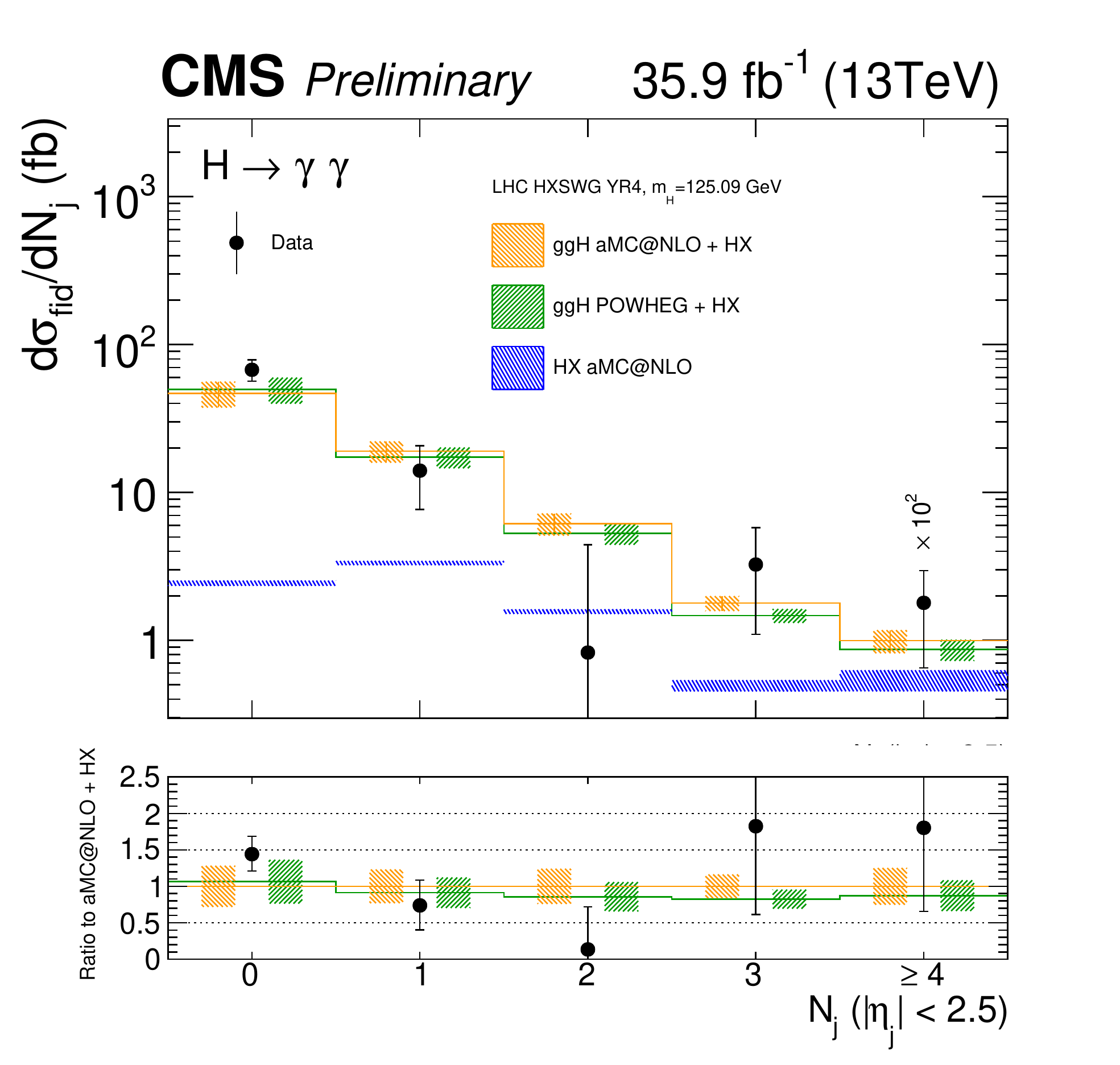}
\caption{
  (Top left) ATLAS diphoton invariant mass spectrum~\cite{ATLAS-gg}.
  (Top center and right) ATLAS differential fiducial cross sections~\cite{ATLAS-gg}, for the
  transverse momentum of the Higgs boson and the number of jets.
  (Bottom left) CMS diphoton invariant mass spectrum~\cite{CMS:HIG-16-040}.
  (Bottom center and right) CMS differential fiducial cross sections, for the
  transverse momentum of the Higgs boson and the number of jets~\cite{CMS:HIG-17-015}.
}
\label{fig:figure-gg}
\end{figure}

\begin{figure}[htb]
\centering
\includegraphics[height=2in]{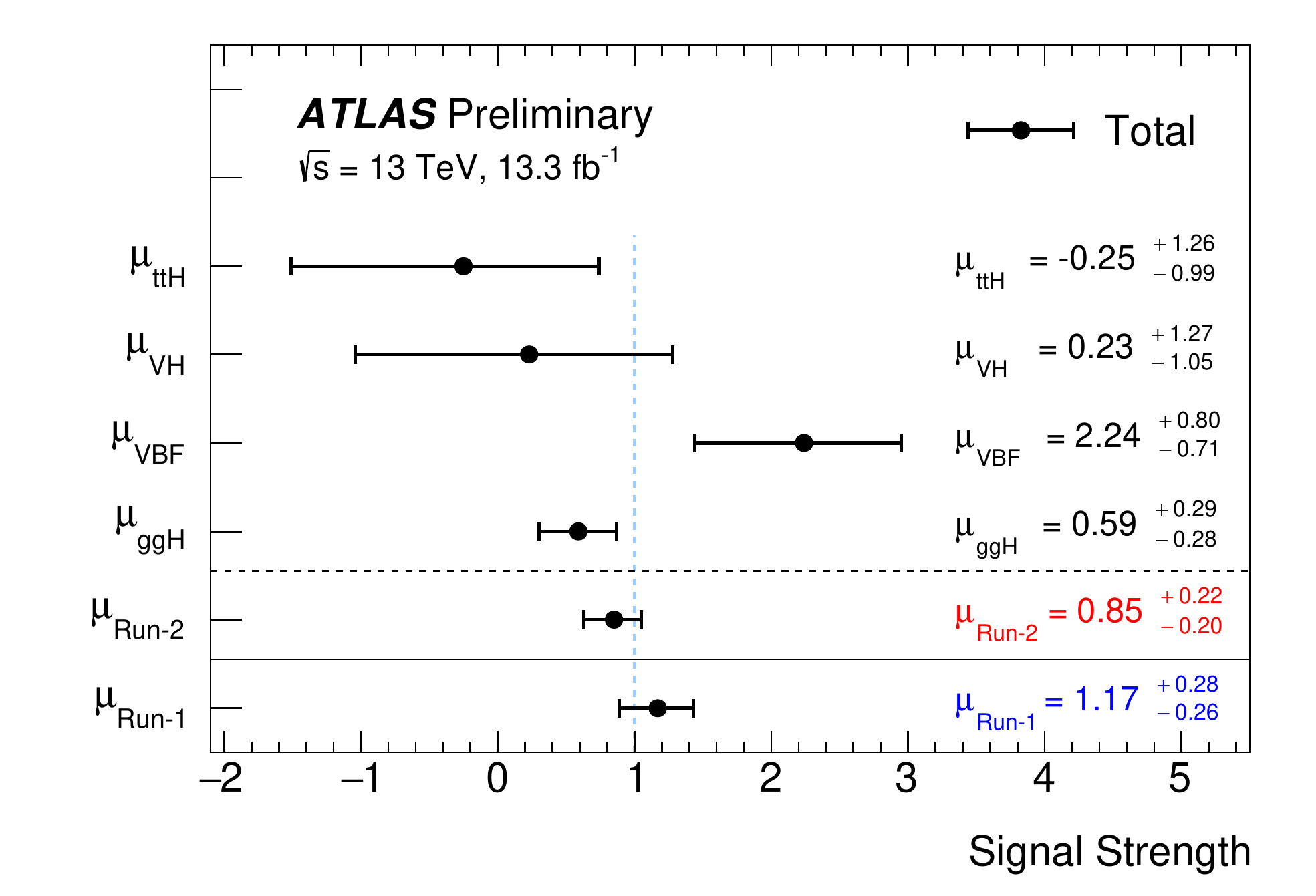}
\includegraphics[height=2in]{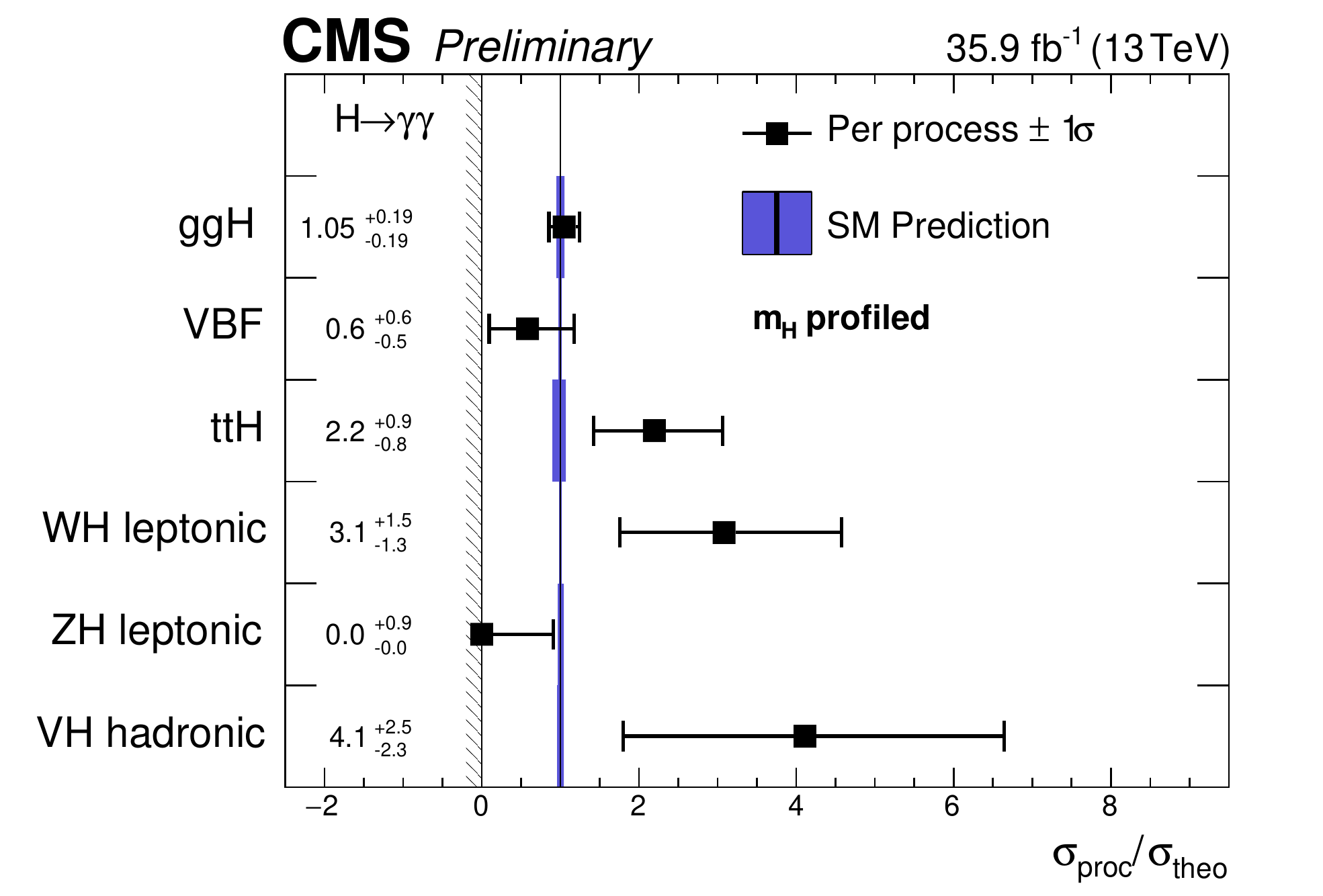}
\caption{
  (Left) The ATLAS ${\rm pp\to H\to \gamma\gamma}$ signal strength measured for
  the different production processes (ggH, VBF, VH and ${\rm ttH}$) and
  globally, compared to the global signal strength measured at 7 and 8~TeV~\cite{ATLAS-gg}.
  The error bars show the total uncertainty.
  (Right) The CMS ${\rm pp\to H\to \gamma\gamma}$ cross section ratios measured
  for each process (black squares) in the Higgs boson simplified template cross
  section framework~\cite{CMS:HIG-16-040}, for profiled ${\rm m_H}$, compared to
  the SM expectation and its uncertainties (blue band). The signal strength
  modifiers are constrained to be non-negative, as indicated by the vertical
  line and hashed pattern at zero.
}
\label{fig:figure-gg-2}
\end{figure}

In the ATLAS analysis~\cite{ATLAS-gg} the Higgs boson signal is measured through a
maximum-likelihood fit to the diphoton invariant mass spectrum in the range
${\rm 105 < m_{\gamma\gamma} < 160~GeV}$ for several fiducial regions, for
each bin of the differential distributions, and for each event category used
in extracting the production cross sections and signal strengths. The analysis
is performed on ${\rm 13.3~fb^{-1}}$ at ${\rm \sqrt{s} = 13~TeV}$, and the
diphoton invariant mass for the inclusive data sample can be seen in
Figure~\ref{fig:figure-gg} (top left). These events are split into exclusive
categories that are optimised for the best separation of the Higgs boson
production processes. These categories are ttH, VH, VBF, and untagged for the
remaining events. This last category contains mostly events produced through
gluon fusion. Several results are extracted from this analysis, starting with
a fiducial cross section for ${\rm pp\to H\to \gamma\gamma}$, measured to be
${\rm \sigma_{fid} = 43.2 \pm 14.9(stat.) \pm 4.9(syst.)~fb}$, which is to be
compared with the SM prediction for inclusive Higgs boson production of
${\rm 62.8_{-4.4}^{+3.4}~fb}$. The gluon fusion contribution to the SM
prediction is N3LO. The differential cross sections for ${\rm pp\to H\to \gamma\gamma}$
as a function of the diphoton transverse momentum and number of jets are shown
in Figure~\ref{fig:figure-gg} (top center and top right). The production mode
event categories are used to determine simplified template cross sections and
total production mode cross sections, as well as the corresponding signal
strengths. In these fits, the cross sections of the ${\rm bbH}$ and tH
production processes are fixed to the expected values from the SM. With the
present dataset, the observed significance of the ${\rm H\to\gamma\gamma}$ signal
is $4.7\sigma$, while $5.4\sigma$ is expected for a SM Higgs boson. The corresponding
signal strengths measured for the different production processes, and globally,
are summarised in Figure~\ref{fig:figure-gg-2} (Left), which also shows the global
signal strength measured in Run-I.

In~\cite{CMS:HIG-17-015} and~\cite{CMS:HIG-16-040} CMS has measured the
integrated and differential fiducial production cross sections for the Higgs
boson in the diphoton decay channel, using ${\rm 35.9~fb^{-1}}$ of proton-proton
collision data collected at ${\rm \sqrt{s} = 13~TeV}$. In
Figure~\ref{fig:figure-gg} (bottom left) the diphoton invariant mass is shown,
and in Figure~\ref{fig:figure-gg} (bottom center and right) the measurement of
the differential cross section is reported as a function of the Higgs boson
transverse momentum and the jet multiplicity. The fiducial cross
section is measured to be ${\rm \sigma_{fid} = 84\pm 11(stat.)\pm 7(syst.)~fb}$, to
be compared with the SM prediction of ${\rm 75 \pm 4~fb}$. This measurement is the
most precise to date. We also report the overall signal strength, the rates for
signal strengths ${\rm \mu_{VBF,VH}}$ and ${\rm \mu_{ggH,ttH}}$ for the model with
these two parameters allowed to vary, and cross section ratios for the Stage 0 STXS
process definitions, as shown in Figure~\ref{fig:figure-gg-2} (right). The best fit
signal strength obtained profiling ${\rm m_H}$ is reported to be
${\rm \mu = 1.16_{-0.10}^{+0.11}(stat.)_{-0.08}^{+0.09}(syst.)_{-0.05}^{+0.06}(theo.)}$.
The best-fit values for the signal strength modifiers associated with the ggH and
${\rm ttH}$ production mechanisms, and with the VBF and VH production processes
are measured; the best fit values for each modifier are
${\rm \mu_{ggH,ttH} = 1.19_{-0.18}^{+0.20}}$ and
${\rm \mu_{VBF,VH} = 1.01_{-0.51}^{+0.57}}$. When ${\rm \mu_{ttH}}$ is considered
separately, the best-fit value is ${\rm \mu_{ttH} = 2.2_{-0.8}^{+0.9}}$, corresponding
to a $3.3\sigma$ excess with respect to the absence of ${\rm \mu_{ttH}}$ production,
and compatible within $1.6\sigma$ with the SM ${\rm \mu_{ttH}}$ prediction.

\section{\boldmath ${\rm H}\to\tau\tau$}

To establish the mass generation mechanism for fermions, it is necessary to
demonstrate the direct coupling of the scalar boson to fermions, and the
proportionality of its strength to the fermion mass. The most promising decay
channel is $\tau\tau$, because of the large event rate expected in the SM
compared to the other leptonic decay modes, and of the smaller contribution
from background events with respect to the ${\rm b\bar{b}}$ channel. Here we
report the results of a search for the SM scalar boson using $35.9~{\rm fb^{-1}}$
at 13 TeV, when it decays to a pair of $\tau$ leptons~\cite{Sirunyan:2017khh}. The
four $\tau$-pair final states with the largest branching fractions,
$\mu\tau_{\rm h}$, ${\rm e}\tau_{\rm h}$, $\tau_{\rm h}\tau_{\rm h}$, and
${\rm e}\mu$, are studied.

\begin{figure}[htb]
\centering
\includegraphics[height=2in]{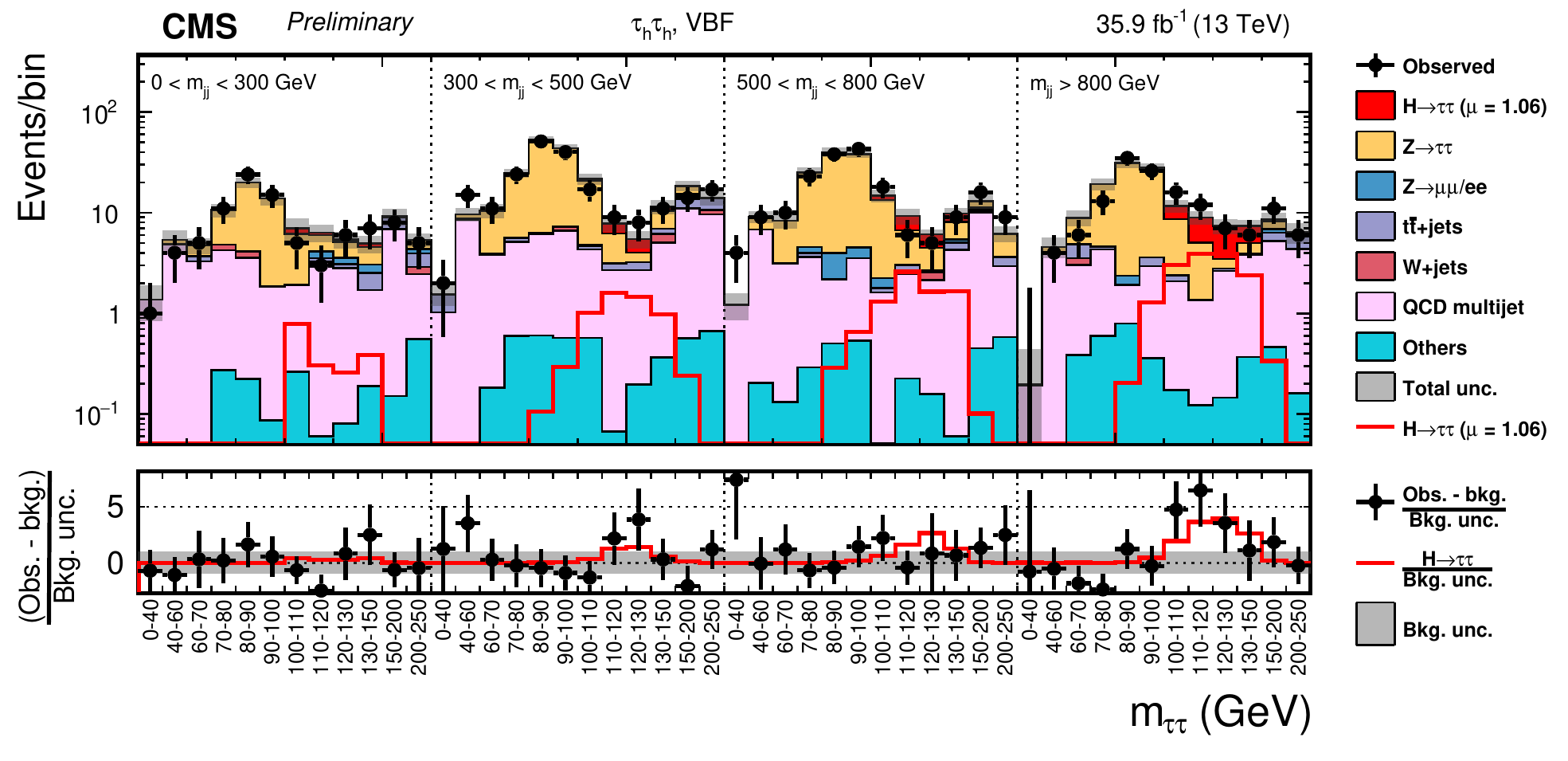}
\caption{
  Observed and predicted 2D distributions in the VBF category of the
  $\tau_{\rm h}\tau_{\rm h}$ final state~\cite{Sirunyan:2017khh}. The normalization
  of the predicted background distributions corresponds to the result of the
  global fit. The signal distribution is normalized to its best-fit signal
  strength.
}
\label{fig:tauhtauh_VBF}
\end{figure}

The search for an excess of SM scalar boson events over the expected background
involves a global maximum likelihood fit based on two-dimensional distributions
in all channels, together with control regions for the ${\rm t{\bar t}}$, QCD
multijet and W+jets backgrounds. Figure~\ref{fig:tauhtauh_VBF} shows the
distribution observed, together with the expected background and signal
distributions, in the $\tau_{\rm h}\tau_{\rm h}$ channel and VBF category. The
signal prediction for a scalar boson with $m_{\rm H} = 125$~GeV is
normalized to its best-fit cross section times branching fraction. The background
distributions are adjusted to the results of the global maximum likelihood fit.
The best fit of the product of the observed ${\rm H\to\tau\tau}$ signal
production cross section and branching fraction is $1.09_{-0.26}^{+0.27}$ times
the standard model expectation. The combination with the corresponding
measurement performed with data collected by the CMS experiment at
center-of-mass energies of 7~TeV and 8~TeV leads to an observed significance of
5.9 standard deviations, equal to the expected significance. This is the first
observation of Higgs boson decays to $\tau$ leptons by a single experiment.

\section{Searches for new phenomena}

Many searches for dark matter (DM) at the LHC involve missing transverse momentum
produced in association to detectable particles. Here we present an updated search
by the ATLAS experiment~\cite{Aaboud:2017uak} for DM associated with the SM
Higgs boson decaying to a pair of photons using $36.1~{\rm fb^{-1}}$ of pp
collision data collected at 13~TeV. Three theoretical benchmark models are
considered in this analysis. In the ${\rm Z_{B}'}$ model a massive vector
mediator ${\rm Z'}$ emits a Higgs boson and subsequently decays to a pair of Dirac
fermionic DM candidates.
The ${\rm Z'}$-${\rm 2HDM}$ model involves the ${\rm Z'}$ boson decaying
to the Higgs boson and an intermediate heavy pseudoscalar boson ${\rm A^{0}}$,
which then decays to a pair of Dirac fermionic DM particles. The third model,
referred to as the heavy scalar model, introduces a heavy scalar boson H
produced primarily via gluon-gluon fusion. Here an effective quartic coupling
between the SM Higgs h, H and two DM particles is considered, with the DM
particle assumed to be scalar. The events that pass a common selection requiring
at least two photon candidates are divided into five categories based on the
event kinematics. These categories have been optimized based either on the ${\rm Z_{B}'}$
and ${\rm Z'}$-${\rm 2HDM}$ signal samples, or using simulated heavy scalar
boson samples, to cover the different kinematic regimes of the heavy scalar
model. The results of the analysis are derived from a likelihood fit of the
$m_{\gamma\gamma}$ distribution in the range of
$105~{\rm GeV} < m_{\gamma\gamma} < 160~{\rm GeV}$.
No significant excess over
the background expectation is observed and 95\% confidence level limits are set
on the production cross section times branching fraction of the SM Higgs boson
decaying into two photons in association with missing transverse energy in the
three different theoretical benchmark models. 95\% confidence level limits are
also set on the observed signal strength in a two-dimensional
$m_{\chi}$-$m_{\rm Z_{B}'}$ plane for the ${\rm Z_{B}'}$ model, and the
$m_{\rm A^0}$-$m_{\rm Z'}$ plane for the ${\rm Z'}$-${\rm 2HDM}$ model. In the
model involving heavy scalar production, 95\% confidence level upper limits are
set on the production cross section times the branching fraction of
${\rm H}\to{\rm h}\chi\chi$, for a dark matter
particle with mass of 60 GeV. The heavy scalar model is excluded for all the
benchmark points investigated.

Another search for DM has been carried out in~\cite{ATLAS-bb-BSM}. In this case
DM is searched for in association with a SM-like Higgs boson decaying to a pair
of b-quarks, using $36.1~{\rm fb^{-1}}$ of pp collisions at 13 TeV recorded with
the ATLAS detector. The ${\rm Z'}$-${\rm 2HDM}$ model has been used for the
optimization of the search and its interpretation. Multivariate algorithms are
used to identify jets containing b-hadrons that are expected in
${\rm h}\to{\rm b{\bar b}}$ decays.




\begin{thebibliography}{99}



\bibitem{Aad:2012tfa} 
  G.~Aad {\it et al.}  [ATLAS Collaboration],
  Phys.\ Lett.\ B {\bf 716}, 1 (2012)
  [arXiv:1207.7214 [hep-ex]].
  
  
\bibitem{Chatrchyan:2012ufa} 
  S.~Chatrchyan {\it et al.}  [CMS Collaboration],
  Phys.\ Lett.\ B {\bf 716}, 30 (2012)
  [arXiv:1207.7235 [hep-ex]].


\bibitem{Chatrchyan:2013lba}
  S.~Chatrchyan {\it et al.} [CMS Collaboration],
  JHEP {\bf 1306} (2013) 081
  [arXiv:1303.4571 [hep-ex]].


\bibitem{Aad:2008zzm} 
  G.~Aad {\it et al.} [ATLAS Collaboration],
  JINST {\bf 3}, S08003 (2008).


\bibitem{Chatrchyan:2008aa} 
  S.~Chatrchyan {\it et al.} [CMS Collaboration],
  JINST {\bf 3}, S08004 (2008).


\bibitem{Aaboud:2017oem} 
  M.~Aaboud {\it et al.} [ATLAS Collaboration],
  Submitted to JHEP
  [arXiv:1708.02810 [hep-ex]].


\bibitem{Sirunyan:2017exp} 
  A.~M.~Sirunyan {\it et al.} [CMS Collaboration],
  Submitted to JHEP
  [arXiv:1706.09936 [hep-ex]].


\bibitem{ATLAS-gg}
  G.~Aad {\it et al.}  [ATLAS Collaboration],
  https://cds.cern.ch/record/2206210
  ATLAS-CONF-2016-067.


\bibitem{CMS:HIG-17-015}
  CMS Collaboration  [CMS Collaboration],
  https://cds.cern.ch/record/2257530
  CMS-PAS-HIG-17-015.


\bibitem{CMS:HIG-16-040}
  CMS Collaboration  [CMS Collaboration],
  https://cds.cern.ch/record/2264515
  CMS-PAS-HIG-16-040.


\bibitem{Sirunyan:2017khh} 
  A.~M.~Sirunyan {\it et al.} [CMS Collaboration],
  Submitted to Phys.\ Lett.\ B
  [arXiv:1708.00373 [hep-ex]].


\bibitem{Aaboud:2017uak} 
  M.~Aaboud {\it et al.} [ATLAS Collaboration],
  Submitted to Phys.\ Rev.\ D
  [arXiv:1706.03948 [hep-ex]].


\bibitem{ATLAS-bb-BSM}
  G.~Aad {\it et al.}  [ATLAS Collaboration],
  http://cds.cern.ch/record/2259066
  ATLAS-CONF-2017-028.


\end{thebibliography}
\end{document}